\documentclass[sigconf]{acmart}
\renewcommand\footnotetextcopyrightpermission[1]{}
\usepackage{graphicx}
\usepackage{amsmath}
\usepackage{booktabs}
\usepackage{amsfonts}
\usepackage{multirow}
\usepackage{makecell}
\usepackage{subfig}
\usepackage{color}
\usepackage{bm}
\usepackage{epstopdf}
\usepackage{url}
\usepackage[cal=cm]{mathalfa}
\usepackage{balance}
\usepackage{threeparttable}
\usepackage{wrapfig}
\usepackage{tabularx}
\usepackage{array}
\usepackage{enumitem}
\usepackage{appendix}
\usepackage{tikz}
\usepackage{float} 
\usepackage[table]{xcolor}
\usepackage{array}
\usepackage[linesnumbered,ruled,vlined]{algorithm2e}
\usepackage[most]{tcolorbox}

\SetKwProg{Stage}{Stage}{}{} 

\setlist[itemize]{leftmargin=*}

\AtBeginDocument{}

\title{SARM: LLM-Augmented Semantic Anchor for \\End-to-End Live-Streaming Ranking}

\setcopyright{acmcopyright}
\copyrightyear{2026}
\acmYear{2026}
\acmDOI{XXXXXXX}

\acmConference[Conference acronym 'XX]{Make sure to enter the correct
  conference title from your rights confirmation email}{June 03--05,
  2018}{Woodstock, NY}

\begin{document}

\author{
    Ruochen Yang\textsuperscript{1,2,3 \dag *}, 
    Yueyang Liu\textsuperscript{1 \dag},
    Zijie Zhuang\textsuperscript{1 \dag},
    Changxin Lao\textsuperscript{1},
    Yuhui Zhang\textsuperscript{1},
    Jiangxia Cao\textsuperscript{1 \ddag}, \\
    Jia Xu\textsuperscript{1}, 
    Xiang Chen\textsuperscript{1},
    Haoke Xiao\textsuperscript{1},
    Xiangyu Wu\textsuperscript{1},
    Xiaoyou Zhou\textsuperscript{1},
    Xiao Lv\textsuperscript{1}, \\
    Shuang Yang\textsuperscript{1}, 
    Tingwen Liu\textsuperscript{2,3 \ddag},
    Zhaojie Liu\textsuperscript{1},
    Han Li\textsuperscript{1},
    Kun Gai\textsuperscript{1}
}
\affiliation{
\institution{
    \textsuperscript{1}Kuaishou Technology, Beijing, China \\
    \textsuperscript{2}Institute of Information Engineering, Chinese Academy of Sciences, Beijing, China \\
    \textsuperscript{3}School of Cyber Security, University of Chinese Academy of Sciences, Beijing, China}
\country{
    \{yangruochen, liutingwen\}@iie.ac.cn, 
    \{liuyueyang05, zhuangzijie, laochangxin, zhangyuhui06, caojiangxia, \\ xujia10, chenxiang08, xiaohaoke, wuxiangyu05, zhouxiaoyou, lvxiao03, \\ yangshuang08, zhaotianxing, lihan08\}@kuaishou.com},
    gai.kun@qq.com
}
\thanks{
    * Work done during an internship at Kuaishou Technology. \\
    \dag \ Equal Contribution. \\
    \ddag \ Corresponding Authors.
}
\renewcommand{\shortauthors}{Ruochen Yang, Yueyang Liu, Zijie Zhuang et al.}

\begin{abstract}

Large-scale live-streaming recommendation requires precise modeling of non-stationary content semantics under strict real-time serving constraints. 
In industrial deployment, two common approaches exhibit fundamental limitations: discrete semantic abstractions sacrifice descriptive precision through clustering, while dense multimodal embeddings are extracted independently and remain weakly aligned with ranking optimization, limiting fine-grained content-aware ranking.
To address these limitations, we propose \textbf{SARM}, an end-to-end ranking architecture that integrates natural-language semantic anchors directly into ranking optimization, enabling fine-grained author representations conditioned on multimodal content.
Each semantic anchor is represented as learnable text tokens jointly optimized with ranking features, allowing the model to adapt content descriptions to ranking objectives.
A lightweight dual-token gated design captures domain-specific live-streaming semantics, while an asymmetric deployment strategy preserves low-latency online training and serving.
Extensive offline evaluation and large-scale A/B tests show consistent improvements over production baselines. 
SARM is fully deployed and serves over 400 million users daily.

\end{abstract}




\maketitle

\section{Introduction}

\begin{figure}[t!]
\begin{center}
\includegraphics[width=8.5cm]{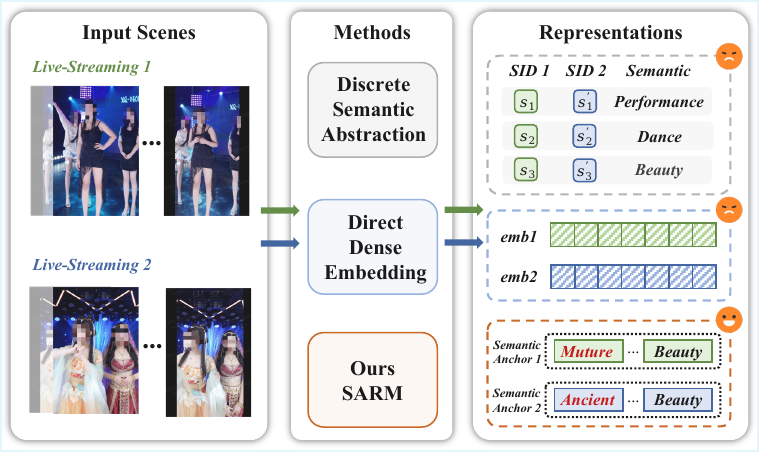}
% \vspace{-0.2cm}
\caption{Comparison of semantic representations: (Top) coarse discrete semantics, (Middle) dense embeddings misaligned with ranking objectives, and (Bottom) fine-grained ranking-aware Semantic Anchors.}
\label{fig:motivation}
\end{center}
\vspace{-0.5cm}
\end{figure}

Live-streaming~\cite{moment, farm} has become a dominant interactive media where ranking systems must interpret non-stationary multimodal content under strict real-time serving constraints.
At industrial scale, user engagement is tightly coupled with short-lived semantic signals generated during streams, making content understanding a core bottleneck for live-streaming recommendation. 
Unlike relatively static scenarios such as short-video~\cite{pepnet, chen2024multi} or e-commerce~\cite{dien} recommendation, live-streaming ranking requires continual semantic adaptation under tight latency and capacity budgets.

Current industrial live-streaming recommendation systems still rely heavily on shallow signals such as titles, author identifiers, and coarse content categories~\cite{mmbee, moment, contentctr}. 
These low-dimensional signals lack the contextual richness to capture spoken language, visual dynamics, and interaction patterns. 
Therefore, models overfit historical behaviors rather than performing content-aware reasoning, limiting generalization and degrading cold-start performance.

As shown in Figure \ref{fig:motivation}, to inject richer semantics, recent work leverages MLLMs to inject multimodal understanding into recommendation, primarily following two paradigms:

\begin{itemize}
\item \textbf{Discrete semantic abstractions} aggregate content into tags or semantic IDs (SIDs)~\cite{larm, foresight}, offering interpretability and inductive bias. 
However, compressing rich multimodal signals into a finite semantic codebook limits representational capacity and weakens fine-grained personalization.
\item \textbf{Dense multimodal embeddings} preserve high-dimensional semantics~\cite{wu2025muse, lemur} but suffer from limited semantic interpretability and cross-modal misalignment, complicating optimization.
\end{itemize}

In production pipelines, both paradigms introduce intermediate semantic layers that decouple content understanding from ranking optimization.
Although SIDs can be optimized end-to-end, compressing rich multimodal content into a finite vocabulary introduces an unavoidable information bottleneck, limiting how much fine-grained semantics can reach the ranking model.
Meanwhile, directly deploying MLLMs or raw multimodal inputs is impractical in live-streaming systems due to strict latency and resource constraints.
These limitations reveal a core deployment trade-off: models must retain rich multimodal semantics while remaining compact enough to support real-time inference and streaming learning under tight industrial capacity constraints.

To address this trade-off, we redesign the live-streaming ranking pipeline to treat semantic anchors as first-class ranking units, directly coupling content understanding with ranking optimization under industrial serving constraints.
We propose SARM, a unified ranking framework that bridges expressive content understanding and industrial deployment requirements. 
SARM consists of three key components: Semantic Anchors, Semantic Anchor Encoder (SAE), and an end-to-end ranking model that unifies semantic anchors with conventional ranking features. 
\textbf{First}, Semantic Anchors encode each live stream through a natural-language description generated offline by a pretrained MLLM using only stream content available at the time.
Their token embeddings are directly optimized within the ranking objective, so semantic anchors function as trainable ranking units rather than frozen textual features.
Unlike tags or SIDs, semantic anchors are not restricted by a fixed vocabulary, avoiding discrete bottlenecks while preserving fine-grained semantics.
\textbf{Second}, to efficiently exploit semantic anchors online, SAE introduces a dual-token gated fusion design tailored to live-streaming domains, where specialized token vocabularies act as a structured lookup table~\cite{cheng2026engram} that encodes streaming terminology and IP-level entities directly at the embedding layer. 
This dual-token design provides explicit grounding for domain-specific live-streaming terminology while remaining fully compatible with future open-vocabulary semantic anchors, ensuring extensibility rather than locking the system into a fixed token space.
\textbf{Finally}, the end-to-end ranking model seamlessly integrates anchors into existing industrial ranking pipelines, enabling joint optimization with conventional features under standard objectives.

We summarize our main contributions as follows:
\begin{itemize}
    \item We introduce Semantic Anchors, natural-language representations generated offline by a pretrained MLLM that preserve rich multimodal semantics without discrete compression.
    \item We propose Semantic Anchor Encoder, a lightweight dual-token framework that enables efficient online learning over semantic anchors through structured lookup and gated fusion.
    \item We develop SARM, an industrial ranking architecture that unifies semantic anchors with compact text encoders and conventional ranking models for end-to-end multimodal recommendation.
    \item Extensive experiments demonstrate consistent improvements in online engagement and commercial metrics over existing pipelines.
    SARM is deployed on the Kuaishou platform, serving over 400 million users in long-term production A/B tests.
\end{itemize}

\section{Related Work}

\subsection{Multimodal Models}

In recent years, progress in multimodal models has substantially improved the ability to jointly model visual, acoustic, and textual information, enabling the development of high-performance multimodal recommender systems.
In the natural-language domain, Transformer-based models such as BERT~\cite{devlin2019bert}, GPT~\cite{brown2020gpt3}, Qwen~\cite{yang2025qwen3}, and GLM~\cite{zeng2025glm} have been widely adopted to extract semantic representations from textual inputs, including item titles, user comments, and conversational content.
For visual and multimodal understanding, models such as ViT~\cite{dosovitskiy2020vit}, CLIP~\cite{radford2021clip}, BLIP~\cite{li2023blip2}, and Qwen-VL~\cite{Qwen3-VL} are capable of learning fine-grained cross-modal representations from visual content.
In addition, audio and speech models, e.g., Qwen-Audio~\cite{chu2023qwenaudio}, have demonstrated strong performance in speech recognition and audio content understanding, further enriching multimodal representations for recommendation systems.

\subsection{Multimodal Recommendation}

Recent advances have significantly promoted the development of multimodal recommendation systems~\cite{liu2024alignrec, kang2017visually, wei2019mmgcn, yuan2023go, zhou2023bootstrap}.
Existing approaches mainly incorporate multimodal information into recommendation models following two representative paradigms.

\textbf{Discrete semantic abstractions} compress rich multimodal content into compact, human-interpretable identifiers such as tags or Semantic IDs (SIDs)~\cite{larm, foresight, saviorrec, eager, qarm, xu2025mmq}, providing inductive bias and improving controllability.
Representative SID-based approaches~\cite{rajput2023tiger, zheng2024lcrec, deng2025onerec} tokenize item semantics into sequences of discrete codes via RQ-VAE or RQ-Kmeans and learn their embeddings within the recommendation objective.
Despite their interpretability, abstracting continuous multimodal signals into a finite codebook can create a discrete bottleneck, limiting representational capacity and weakening fine-grained personalization.

\textbf{Dense multimodal embeddings} instead rely on pretrained multimodal encoders to produce continuous semantic representations from images and texts, which are then fed into downstream recommenders~\cite{liu2024alignrec, wang2023missrec, wu2025muse, lemur}.
AlignRec~\cite{liu2024alignrec} leverages collaborative filtering supervision to align CLIP-based multimodal representations with item ID embeddings, enabling semantic information to be injected into ID-based recommendation frameworks.
MISSRec~\cite{wang2023missrec} introduces trainable adapter modules to transform general-purpose pretrained multimodal features into representations that are more suitable for personalized recommendation tasks, while keeping the backbone multimodal encoders fixed.
Similarly, MUSE~\cite{wu2025muse} incorporates pretrained multimodal features into both the General Search Unit and the Exact Search Unit, leveraging semantic similarity signals to enhance lifelong user interest modeling without jointly training the multimodal encoders with the recommendation objective.
While preserving high-dimensional semantics, dense embeddings often lack explicit semantic interpretability and may suffer from cross-modal misalignment, which can complicate downstream optimization.

\section{Methodology}

\begin{figure*}[t!]
% \vspace{-0.1cm}
\begin{center}
\includegraphics[width=17.0cm]{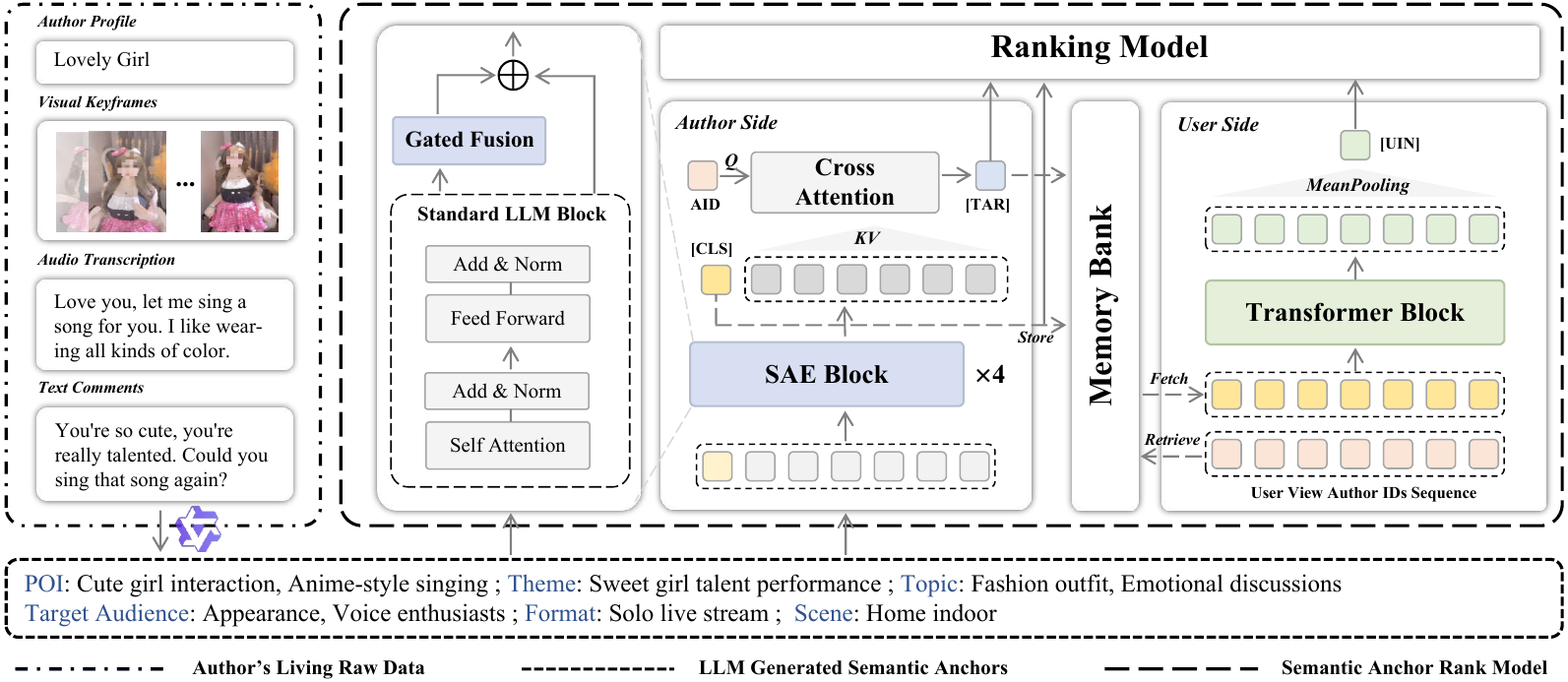}
% \vspace{-0.3cm}
\caption{
Overall architecture of SARM. 
A fine-tuned MLLM generates semantic anchors that are encoded and optimized end-to-end with ranking objectives. 
A memory bank maintains representations for efficient online training and inference.
}
\vspace{-0.5cm}
\label{fig:model}
\end{center}
\end{figure*}

In this section, we present the architecture of SARM. 
The overall framework is illustrated in Figure \ref{fig:model}.
We first describe Semantic Anchors, a ranking-aware semantic representation of live-streaming content implemented as a sequence of natural-language tokens.
We then introduce the Semantic Anchor Encoder (SAE), a lightweight semantic encoder tailored to live-streaming scenarios and trained end-to-end within the ranking pipeline. 
SARM connects SAE with the ranking model into a unified trainable architecture. 
Finally, we describe our asymmetric optimization and deployment strategy, designed to balance real-time training and inference under practical online constraints.

\subsection{Semantic Anchor}

We encode each live-streaming into a structured natural-language description, termed semantic anchor. It acts as a persistent interface bridging multimodal content understanding and ranking.

\subsubsection{\textbf{Anchor Representation}}
Semantic Anchors are generated offline by a fine-tuned multimodal large language model (MLLM)~\cite{qwen2.5vl, Qwen3-VL} using a fixed prompt template that summarizes visual, acoustic, and contextual attributes of 
an author into natural-language textual form.
The format of a semantic anchor is illustrated blow:
\begin{tcolorbox}[
    float=h!,
    title=The Format of Semantic Anchor,
    width=\linewidth, 
    colframe=gray!60
]
    \small 
    \textcolor[HTML]{2F528F}{[CLS]} POI: phrase1, phrase2, \dots \\
    \textcolor[HTML]{2F528F}{[SEP]} Theme: phrase1, phrase2, \dots \\
    \textcolor[HTML]{2F528F}{[SEP]} Topic: phrase1, phrase2, \dots \\
    \textcolor[HTML]{2F528F}{[SEP]} Target audience: phrase1, phrase2, \dots \\
    \textcolor[HTML]{2F528F}{[SEP]} Format: phrase1, phrase2, \dots \\
    \textcolor[HTML]{2F528F}{[SEP]} Scene: phrase1, phrase2 \\
    \textcolor[HTML]{2F528F}{[PAD] ... [PAD]}
    \end{tcolorbox}
\vspace{-0.3cm}

Formally, \textbf{given a live author $\mathbf{s}$, its corresponding semantic anchor is a token sequence:}
\begin{equation}
    A_{s} = \{t_1, t_2, ...,t_n\},
\end{equation}
where each token $t_i$ is drawn from the standard vocabulary of the underlying language model and jointly forms a natural-language description of the live-streaming.
Unlike discrete tags or semantic IDs, semantic anchors inherit the open-ended structure of natural-language rather than a predefined semantic vocabulary.
This design enables seamless compatibility with future MLLMs, allowing the system to benefit from richer semantic generation without redesigning the representation space.

\subsubsection{\textbf{Anchor Generation via MLLM}}

To generate semantically consistent anchors across live-streaming sessions, we design a domain-specific multimodal data aggregation framework that integrates author original profile with heterogeneous signals: visual keyframes, audio signals, and user comments. 
Each modality is processed through a pipeline tailored to live-streaming characteristics
followed by domain-specific fine-tuning of a MLLM:

\begin{itemize}
\item \textbf{Visual Keyframes.} Guided by content diversity and user feedback, we dynamically sample around 20 keyframes per stream, prioritizing facial close-ups and representative scenes.

\item \textbf{Audio Transcription.} Automatic speech recognition (ASR) model is applied to denoise and transcribe audio segments within fixed temporal windows aligned with sampled keyframes, extracting domain-relevant spoken content.

\item \textbf{Text Comments.} User comments are filtered based on users' engagement value and semantic relevance, retaining the top 32 representative comments per stream that reflect authentic user interaction patterns
\end{itemize}

The aggregated multimodal signals are fused through a structured prompt template (see Appendix \ref{app:template}) and fed into the MLLM to produce semantic anchors following a consistent schema across six dimensions: Point of Interest (POI), Theme, Topic, Target Audience, Format, and Scene. 
Through systematic data curation and aggregation, we distill a computationally efficient MLLM optimized for online inference in live-streaming environments.

\subsection{Semantic Anchor Encoder}

While semantic anchors provide rich natural-language representations, directly coupling a large LLM-style text encoder with ranking models suffers from two fundamental limitations.
First, encoding long descriptions with large language models is incompatible with real-time ranking and online end-to-end training constraints.
Second, live-streaming content contains dense domain terminology and IP-specific entities, such as \textit{Lao Tie} and \textit{PUBG}.
When such terms are split into multiple tokens by standard LLM tokenizers, lightweight language models struggle to recover their underlying semantics.
To address these issues, we introduce the \textbf{Semantic Anchor Encoder (SAE)}.

\subsubsection{\textbf{Live-Streaming Tokenizer.}}
Standard LLM tokenizers are trained on large general-domain corpora and are not optimized for domain-specific expressions in live-streaming settings.
As a result, many live-streaming terms and IP entities are fragmented into multiple subword tokens.
For example, a game entity such as \textit{PUBG} may be tokenized as:

\[
\textit{PUBG}\rightarrow(t_1, t_2, t_3)=('P', 'UB', 'G').
\]
While large models can partially recover such structure, lightweight language encoders lack sufficient capacity to reconstruct the underlying semantics from fragmented tokens.

To address this limitation, we introduce a Live-Streaming Tokenizer that augments the base tokenizer with domain-specific 
terminology expressions.
We collect semantic anchors from multiple days of live-streaming content and apply Byte Pair Encoding (BPE)~\cite{bpe} over sequences produced by the base tokenizer.
Formally, let a semantic anchor be tokenized into a token sequence $A_s = \{t_1, t_2, \dots, t_n\},  t_i \in \mathcal{V}_{\text{ llm}}$ , where \(\mathcal{V}_{\text{llm}}\) denotes the vocabulary of the original tokenizer.  
BPE is then applied iteratively: at each step, the most frequent adjacent pair of tokens (with a threshold of 100 thousand) is merged into a new token $u$:
\begin{equation}
    (t_a, t_b) \;\rightarrow\; u, \qquad u \in \mathcal{V}_{\text{new}}.
\end{equation}

We iteratively apply BPE merging over the anchor corpus to construct a set of domain tokens $\mathcal{V}_{\text{new}}$.
After multiple rounds of merging, we obtain an extended tokenizer adapted to live-streaming scenarios, where frequently co-occurring terms and IP entities are represented as atomic semantic units while all original tokens remain intact.
Note that following the initial acquisition, we continuously persist semantic anchor texts daily and incrementally update the Live-Streaming Tokenizer accordingly.
With the proposed Live-Streaming Tokenizer, the semantic anchor is mapped into an enhanced token sequence $A'_{s}$.

\subsubsection{\textbf{Gated Fusion.}} \label{sec:gated}

\begin{figure}[t!]
% \vspace{-0.1cm}
\begin{center}
\includegraphics[width=8.2cm]{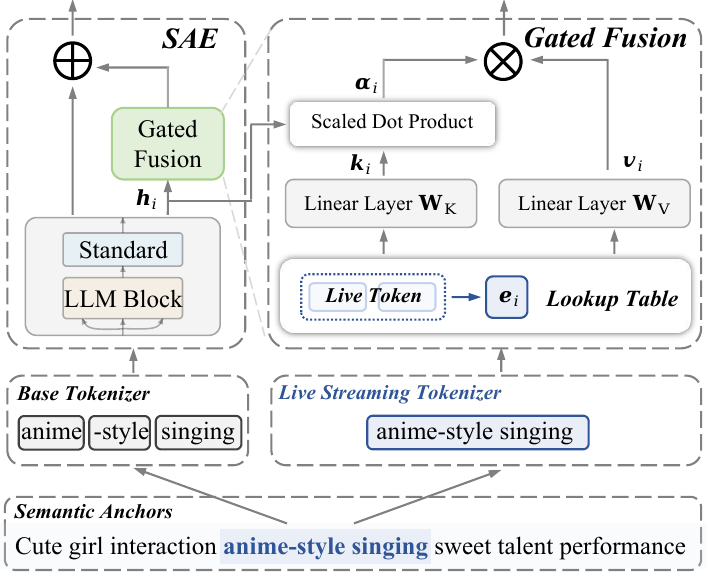}
% \vspace{-0.3cm}
\caption{
Detailed architecture of the gated fusion module. The Live-Streaming Tokenizer aggregates domain-specific terms, and the gated fusion module integrates the resulting embeddings into a standard LLM via external lookup, where an example tokenization process is illustrated for clarity.
}
% \vspace{-0.5cm}
\label{fig:gate_fusion}
\end{center}
\end{figure}

Although extending the tokenizer with domain-specific tokens reduces the learning burden of lightweight encoders, directly replacing the base LLM tokenizer with a Live-Streaming Tokenizer introduces a trade-off between domain specialization and general language capability. 
We therefore require a fusion mechanism that selectively injects domain semantics while preserving the base language model’s general language understanding. 
Such a mechanism should satisfy two properties:
(1) domain semantics are incorporated only when beneficial for ranking;
(2) the fusion remains lightweight and compatible with online training.

Concretely, we propose a gated semantic fusion architecture that preserves the base LLM tokenizer and backbone while injecting domain semantics through an external semantic memory.
As demonstrated in Fig.~\ref{fig:gate_fusion}, instead of replacing the base sequence $A_s$ with $A'_s$ and feeding it directly into a LLM backbone, we treat $A'_s$ as an auxiliary lookup memory produced by the Live-Streaming Tokenizer.
Given a token $t'_i$ from $A'_s$, we compute a token-wise gated interaction between its embedding $e'_i$ and the intermediate hidden state $h_i$ obtained from encoding $A_s$ with the base LLM:

\begin{equation}
\begin{gathered}
\boldsymbol{k}_i = \mathbf{W}_\mathrm{K} \boldsymbol{e}_i,\quad \boldsymbol{v}_i = \mathbf{W}_\mathrm{V} \boldsymbol{e}_i, \\\\
\alpha_i =
\sigma\!\left(
\frac{\text{RMSNorm}(\boldsymbol{h}_i)^{\top}\,\text{RMSNorm}(\boldsymbol{k}_i)}{\sqrt{d}}
\right),
\end{gathered}
\end{equation}
where $\mathbf{W}_K$ and $\mathbf{W}_V$ are learnable projection matrices. 
To fuse the semantics from both tokenizers, we combine the two representations via gated residual addition before passing them to the next transformer layer:
\begin{equation}
\boldsymbol{h}'_i = \boldsymbol{h}_i + \alpha_i \cdot \boldsymbol{v}_i.
\end{equation}

\subsubsection{\textbf{Lightweight Language Model Integration.}} 

To satisfy the strict latency requirements of ranking while enabling end-to-end joint training of token embeddings, the language backbone, and the ranking model, we adopt a lightweight yet expressive BERT-style encoder on the author side, referred to as \textbf{SAE}. 
The encoder consists of four transformer blocks with single-head self-attention. 
We adopt rotary positional encoding (RoPE)~\cite{rope} to provide relative position awareness and stable extrapolation for streaming inputs.

We integrate semantic anchors into SAE through the gated semantic fusion mechanism described above, injecting domain semantics at certain layers (referred to Appendix \ref{app:position}):
\begin{equation}
    \boldsymbol{h}= \mathrm{SAE}(A_s, A'_s).
\end{equation}
After encoding, we extract the embedding of the [CLS] token position as the aggregated semantic representation of the anchor:
\begin{equation}
\begin{gathered}
\boldsymbol{h}_{\text{CLS}} = \boldsymbol{h}[0].
\end{gathered}
\end{equation}

This representation captures rich multimodal semantics distilled through the semantic anchor.
However, semantic content alone is insufficient to model author-specific behavior.
In practice, users exhibit different preferences toward authors with similar content profiles, revealing identity-driven interaction signals beyond textual semantics.
To model this complementary signal, we introduce an explicit author identity embedding and fuse it with $\boldsymbol{h}$ via cross-attention. 
It enables the model to jointly capture semantic content and author-specific personalization, enriching the representation space beyond the capacity of textual anchors alone:
\begin{equation}\
    \label{eq:tar}
    \boldsymbol{h}^{a}_{\mathrm{TAR}} = \text{CrossAttention}(\boldsymbol{h}^{a}_{id}, \boldsymbol{h}, \boldsymbol{h})
\end{equation}
where $\boldsymbol{h}^{a}_{id}$ is a learnable ID embedding corresponding to author $a$.

Given the temporal continuity of live-streaming content, we maintain a memory bank to cache semantic anchor representations for efficient reuse. The memory bank $\mathcal{M}$ stores precomputed author embeddings indexed by author ID:
\begin{equation}
    \mathcal{M}[a] = (\boldsymbol{h}^a_{\text{CLS}}, \boldsymbol{h}^a_{\text{TAR}}),
\end{equation}
where the superscript $a$ denotes different authors.
This design avoids redundant textual encoding and enables constant-time retrieval during ranking.

On the user side, we model historical interests by retrieving the memory bank with sequence of authors from the user’s effective viewing history. 
The retrieved embeddings form a sequence:
\begin{equation}
\label{eq:user}
    \boldsymbol{h}^u = (\boldsymbol{h}_{\text{CLS}}^{a_0}, \boldsymbol{h}_{\text{CLS}}^{a_1}, \dots, \boldsymbol{h}_{\text{CLS}}^{a_m}) \in \mathbb{R}^{m \times d},
\end{equation}
and a self-attention transformer is then applied to obtain the final user interests:
\begin{equation}
\label{eq:user_interest}
    \boldsymbol{h}^{u}_{\text{UIN}} = \text{MeanPooling}\big(\text{Transformer}(\boldsymbol{h}^u)\big).
\end{equation}

\subsection{\textbf{Semantic Anchor Ranking Model}}

We integrate semantic representations from both author and user sides as inputs to the ranking model. 
The feature set includes the semantic anchor representation $\boldsymbol{h}^a_{\text{CLS}}$, the identity-aware author representation $\boldsymbol{h}^a_{\text{TAR}}$, and the aggregated user interest representation $\boldsymbol{h}^u_{\text{UIN}}$, 
along with the existing ranking features from the backbone model, denoted as $\boldsymbol{h}_{\text{rank}}$.
These features are concatenated and fed into the existing multi-task ranking backbone~\cite{home} to model user–author interactions and predict final ranking scores:
\begin{equation}
    \hat{y}^{\mathrm{xtr}} = \mathrm{MultiTask}\bigl(\mathrm{Concat}[\,\boldsymbol{h}^a_{\mathrm{CLS}},\,
    \boldsymbol{h}^a_{\mathrm{TAR}},\,
    \boldsymbol{h}^u_{\mathrm{UIN}},\,
    \boldsymbol{h}_{\mathrm{rank}}\,]\bigr).
\end{equation}

We refer to this end-to-end architecture as the \textbf{Semantic Anchor Ranking Model (SARM)}.
This model is optimized using multiple binary classification objectives corresponding to different engagement signals.
The overall recommendation loss is defined as the sum of task-wise binary cross-entropy losses:
\begin{equation}
    \mathcal{L}_{\mathrm{rec}} = - \sum_{\mathrm{xtr}}^{\mathrm{Tasks}} 
    \Bigl(
        y^{\mathrm{xtr}} \, \log \hat{y}^{\mathrm{xtr}} 
        + 
        (1 - y^{\mathrm{xtr}}) \, \log \bigl(1 - \hat{y}^{\mathrm{xtr}}\bigr)
    \Bigr).
\end{equation}

In practice, jointly training semantic encoding and ranking introduces optimization challenges.
The semantic space used for text understanding does not fully align with the discriminative space required for ranking, especially on the author side where semantic anchors dominate representation learning.
This mismatch slows convergence and destabilizes training.
To stabilize optimization, we introduce an auxiliary CTR prediction task~\cite{dien} operating on the author-side representation. 
A lightweight MLP followed by a softmax layer is used as the auxiliary prediction head:
\begin{equation}
    \hat{y}_\mathrm{aux} = \text{MLP}(\text{Concat}[\boldsymbol{h}^a_{\mathrm{CLS}}, \boldsymbol{h}^a_{\mathrm{TAR}}]),
\end{equation}
and the binary cross-entropy loss is applied:
\begin{equation}
    \mathcal{L}_{\mathrm{aux}} 
    = 
    - y \, \log \hat{y}_{\mathrm{aux}} 
    - 
    (1 - y) \, \log \bigl(1 - \hat{y}_{\mathrm{aux}}\bigr).
\end{equation}

This auxiliary objective provides direct supervision to the semantic representations, which promotes their alignment with ranking task and enhances their generalization.

The final training objective is a weighted combination of the main recommendation loss and the auxiliary loss:
\begin{equation}
    \mathcal{L} = \mathcal{L}_\mathrm{rec} + \lambda \mathcal{L}_\mathrm{aux},
\end{equation}
where $\lambda$ is a tunable hyperparameter controlling the influence of the auxiliary task.

\subsection{Asymmetric Deployment}

\begin{figure}[t!]
% \vspace{-0.1cm}
\begin{center}
\includegraphics[width=8.2cm]{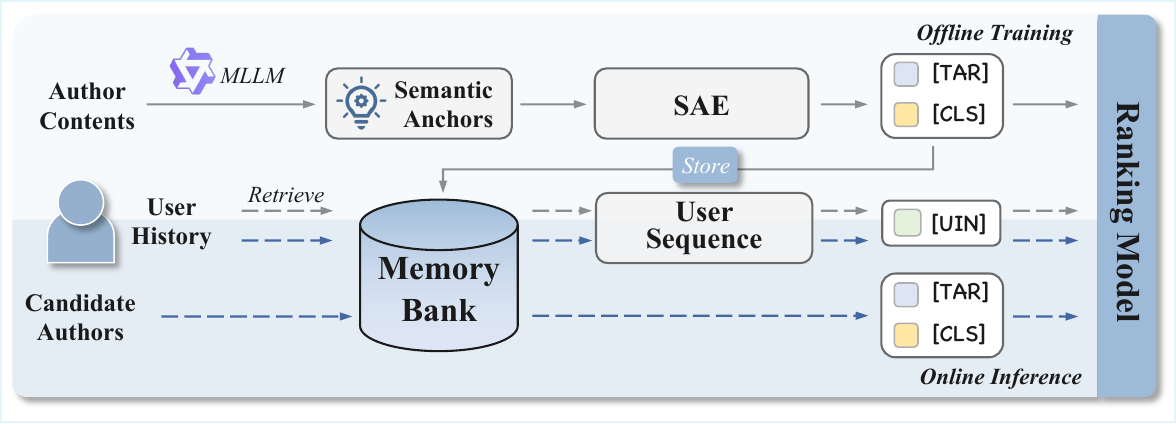}
\caption{Asymmetric deployment pipeline of SARM. 
A memory bank caches author representations to support efficient online training and inference.}
\vspace{-0.5cm}
\label{fig:pipeline}
\end{center}
\end{figure}

Fig.~\ref{fig:pipeline} illustrates the overall pipeline of SARM in the Kuaishou live-streaming recommendation system.
To reduce computational cost and system latency while keeping learned features up to date, we adopt an asymmetric deployment architecture centered around a memory bank.
Specifically, we employ distinct processing strategies for the author side and the user side:

\begin{table*}[t!]
  \caption{
  Offline results for live-streaming ranking. 
  \textit{Bold} indicates the base model, \textit{blue} highlights the best performance. 
  Indentation denotes additional modules built on the preceding configuration.
  }
  % \vspace{-0.3cm}
  % \setlength{\tabcolsep}{7pt}
  \label{tab:offline}
  \begin{tabular}{c|l|cccccccc}
    \toprule
    \multicolumn{1}{c}{\multirow{2.5}{*}{\textbf{Paradigm}}} & \multicolumn{1}{|c|}{\multirow{2.5}{*}{\textbf{Model Variants}}} & \multicolumn{2}{c}{\textbf{CTR}}  & \multicolumn{2}{c}{\textbf{WTR}} & \multicolumn{2}{c}{\textbf{LVTR}} & \multicolumn{2}{c}{\textbf{GTR}} \\
    \cmidrule(r){3-4} \cmidrule(r){5-6} \cmidrule(r){7-8} \cmidrule(r){9-10} & & AUC & GAUC & AUC & GAUC & AUC & GAUC & AUC & GAUC \\
    
    \midrule
    
    - & \textbf{Base Ranking Model} & \textbf{0.8387} & \textbf{0.6453} & \textbf{0.9217} & \textbf{0.6500} & \textbf{0.8928} & \textbf{0.7542} & \textbf{0.9792} & \textbf{0.7319} \\
    
    \midrule
    
    \multirow{3}{*}{\textbf{Two-Stage}} & + Tags & 0.8390 & 0.6457 & 0.9220 & 0.6510 & 0.8932 & 0.7542 & 0.9794 & 0.7324 \\
    & + SIDs (LARM) & 0.8389 & 0.6469 & 0.9221 & 0.6536 & 0.8936 & 0.7553 & 0.9799 & 0.7324 \\
    & + MLLM Emb (MMBee) & 0.8385 & 0.6451 & 0.9210 & 0.6475 & 0.8932 & 0.7551 & 0.9790 & 0.7320 \\
    & + MLLM Emb (SIM) & 0.8394 & 0.6463 & 0.9224 & 0.6509 & 0.8939 & 0.7548 & 0.9792 & 0.7330 \\
    
    \midrule
    
    \multirow{4}{*}{\textbf{End-to-End}} & + SAE (Author Side) & 0.8396 & 0.6475 & 0.9225 & 0.6511 & 0.8948 & 0.7567 & 0.9804 & 0.7342 \\ 
    & \quad + Auxiliary Loss (Author Side) & 0.8398 & 0.6474 & 0.9227 & 0.6513 & 0.8948 & 0.7573 & 0.9803 & 0.7352 \\
    & \quad + User Sequence (User Side) & 0.8405 & 0.6483 & 0.9227 & 0.6519 & 0.8955 & 0.7575 & 0.9816 & 0.7358 \\
    \rowcolor{blue!8}
    \cellcolor{white}\textbf{} & \cellcolor{white}\textbf{+ \textbf{Full (SARM)}} & \textbf{0.8411} & \textbf{0.6485} & \textbf{0.9232} & \textbf{0.6522} & \textbf{0.8959} & \textbf{0.7580} & \textbf{0.9825} & \textbf{0.7369} \\
     
    \bottomrule
  \end{tabular}
  % \vspace{-0.3cm}
\end{table*}

\begin{itemize}
    \item \textbf{Author side (asymmetric design).}
    During training, the target author is used as supervision, and the associated multimodal content is jointly encoded by the MLLM and SAE to produce semantic-aware author representations. In practice, semantic anchor generation via the MLLM is performed daily for each author, while the SAE-based encoding and the end-to-end ranking are conducted through streaming continuous training. The resulting representations are stored in a memory bank indexed by unique author IDs and can be overwritten as new observations accumulate. During inference, given candidate authors retrieved by the upstream recall stage, the system fetches their precomputed representations via constant-time lookup from the memory bank. This design completely bypasses the expensive encoding steps, thereby ensuring efficient online serving.

    \item \textbf{User side (symmetric strategy).}
    In contrast, user interest representations are computed in an identical manner during both training and inference phases. The system retrieves the relevant embeddings from the memory bank based on the user\textquotesingle s historical interaction sequence and encodes them continuously, maintaining modeling consistency across both phases.
\end{itemize}

Finally, multiple related representations are jointly fed into ranking model for optimization and prediction.

\section{Experiments}

In this section, we conduct extensive experiments offline and online to answer the following questions:

\begin{itemize}
    \item RQ1: How does our proposed framework SARM and its corresponding components perform compared to existing methods?
    \item RQ2: How effective and efficient is SARM across various testing environments and testing settings?
    \item RQ3: How does SARM affect online live-streaming services?
\end{itemize}

\subsection{Experimental Settings}

\subsubsection{\textbf{Dataset.}}

To evaluate the adaptability of our model in industrial live-streaming scenarios, we conduct offline experiments on real-world data from the Kuaishou live-streaming platform.
Specifically, our experiment is based on a dataset comprising 400 million users, 3 million authors, and billions of interaction records, derived from a real-time 30-second sliding window data streaming.
The data are chronologically ordered, with the final day used for testing and all preceding days used for training. 
Details of the dataset composition can be found in~\cite{moment}.

\subsubsection{\textbf{Evaluation Metrics.}}

We evaluate model performance using standard XTR prediction metrics, including AUC and GAUC. We consider four engagement objectives: click (CTR), follow (WTR), long view (LVTR), and gift (GTR) as targets in the multi-task setting.

\subsection{Offline Experiments (RQ1)}

\subsubsection{\textbf{Overall Experiments.}}

\begin{figure}[t]
% \vspace{-0.1cm}
\begin{center}
\includegraphics[width=8.3cm]{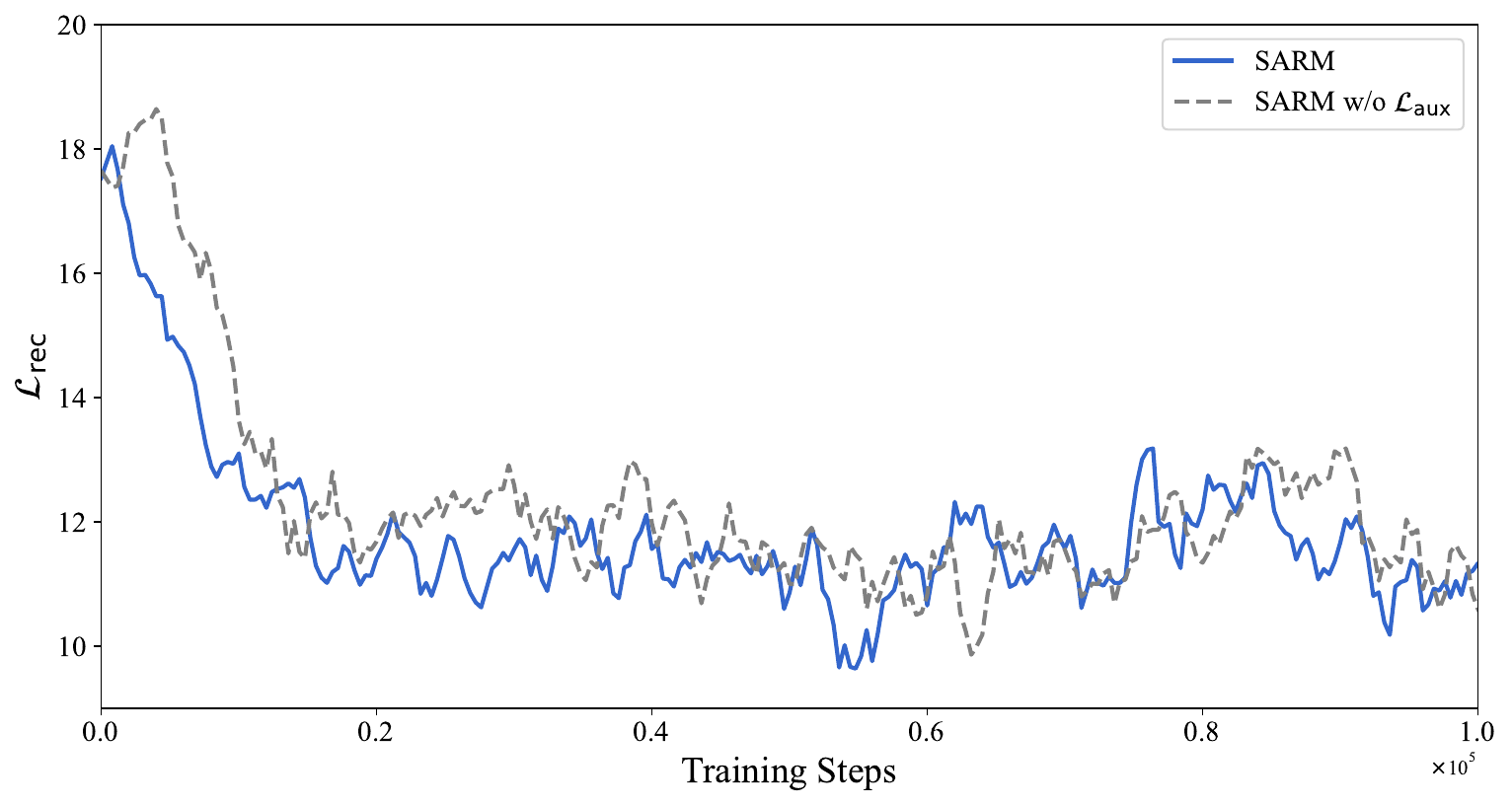}
% \vspace{-0.3cm}
\caption{Effect of the auxiliary loss in training.}
% \vspace{-0.5cm}
\label{fig:loss}
\end{center}
\end{figure}

We conduct comprehensive offline experiments to compare SARM with two-stage multimodal pipelines and to isolate the contribution of its individual components. 
The base ranking model is the deployed HoME~\cite{home}, an industrial MMoE multi-task ranking system incorporating ID embeddings, sequence modeling, and feature crossing.
Unless otherwise stated, all experiments are conducted on this ranking backbone.
Multi-task evaluation results are summarized in Table~\ref{tab:offline}, with detailed baseline descriptions provided in Appendix~\ref{app:baseline}.

We summarize the following observations:
\begin{itemize}
    \item After the base ranking model captures ID-based behavioral signals, incorporating multimodal content features, such as tags, dense embeddings, or semantic anchors, consistently improves performance by providing additional semantic context.
    \item The proposed end-to-end framework SARM outperforms prior two-stage pipelines such as LARM, MMBee, and SIM. 
    It suggests that joint optimization enables tighter alignment between representation learning and ranking objectives, leading to more task-oriented multimodal features and better performance.
    \item Simultaneously modeling semantic information on both the author and user sides proves more effective than single-side modeling.
    It suggests that dual-side representations better align user preferences with author content attributes, leading to more accurate modeling of user-author interactions.
    \item The auxiliary objective supplies dense supervision to the semantic encoder, leading to more stable optimization and faster convergence. 
    Fig.~\ref{fig:loss} also illustrates faster and more stable convergence of the main recommendation loss.
\end{itemize}

\subsubsection{\textbf{Replacement Experiments.}}

\begin{table}[t!]
  \caption{Component replacement in SARM. 
  Each row incrementally adds modules over the previous configuration. 
  \textit{Blue} highlights the best performance within each component.
  }
  % \vspace{-0.3cm}
  \setlength{\tabcolsep}{3pt}
  \label{tab:ablation}
  \begin{tabular}{l|cccc}
    \toprule
    \multicolumn{1}{c|}{\multirow{2.5}{*}{\textbf{Variants}}} & \multicolumn{2}{c}{\textbf{CTR}}  & \multicolumn{2}{c}{\textbf{LVTR}} \\
    \cmidrule(r){2-3} \cmidrule(r){4-5} & AUC & GAUC & AUC & GAUC \\

    \midrule

    \textbf{Base Ranking Model} & \textbf{0.8387} & \textbf{0.6453} & \textbf{0.8928} & \textbf{0.7542} \\
    
    \midrule
    
    \textbf{+ [CLS] (Tokenizer)} & & & & \\
    \quad \textit{w}/ Base Tokenizer & +0.07\% & +0.10\% & +0.08\% & +0.11\% \\
    \quad \textit{w}/ Live-Streaming Tokenizer & +0.06\% & +0.06\% & +0.08\% & +0.10\% \\
    \rowcolor{blue!8}    
    \quad \textit{w}/ \textbf{Gated Fusion} & \textbf{+0.09\%} & \textbf{+0.14\%} & \textbf{+0.12\%} & \textbf{+0.22\%} \\
    
    \midrule

    \textbf{+ [TAR] (Eq.\ref{eq:tar})} & & & & \\
    \quad \textit{w}/ MeanPooling & +0.04\% & +0.07\% & +0.13\% & +0.19\% \\
    \rowcolor{blue!8}
    \quad \textit{w}/ \textbf{Cross Attention} & \textbf{+0.09\%} & \textbf{+0.22\%} & \textbf{+0.20\%} & \textbf{+0.25\%} \\

    \midrule

    \textbf{+ [UIN]  (Eq.\ref{eq:user})} & & & & \\
    \quad \textit{w}/ $\mathrm{[TAR]}$ Sequence & +0.15\% & +0.28\% & +0.15\% & +0.19\% \\
    \rowcolor{blue!8}
    \quad \textit{w}/ \textbf{$\mathrm{[CLS]}$ Sequence} & \textbf{+0.18\%} & \textbf{+0.30\%} & \textbf{+0.27\%} & \textbf{+0.33\%} \\
     
    \bottomrule
  \end{tabular}
  % \vspace{-0.3cm}
\end{table}

We conduct component replacement experiments in an incremental manner, sequentially replacing the modules corresponding to the three newly introduced input features (\textit{i.e.}, [CLS], [TAR], [UIN]) in the SARM ranking model. The results are shown in Table \ref{tab:ablation}.

\textbf{Semantic Anchor Representation [CLS].}
We observe that both the base tokenizer and the Live-Streaming Tokenizer improve ranking performance.
However, directly replacing the base tokenizer with the Live-Streaming Tokenizer does not yield additional gains.
A possible explanation is that merging domain tokens into specialized units weakens their ability to share semantic structure with related tokens through prefix composition, reducing generalization. 
Gated fusion mitigates this trade-off by combining both token spaces, balancing semantic precision and general language compatibility, and achieving the best ranking performance.

\textbf{Identity-Aware Author Representation [TAR].}
We compare two strategies for aggregating the semantic representations, for the first is a simple mean pooling over the SAE outputs $\text{MeanPooling}(h)$, and the second is an identity-aware feature crossover that incorporates the author's personalized profile $h^{a}_{\mathrm{TAR}}$. Results show that introducing the additional author representation leads to a significant improvement in ranking performance, indicating that identity-aware features are critical for distinguishing authors with similar semantic content.

\textbf{Aggregated User Interest Representation [UIN].}
For user interest modeling, we compare sequences constructed from $h^{a}_{\mathrm{CLS}}$ embeddings with those using $h^{a}_{\mathrm{TAR}}$ embeddings. 
Experiments show that $h^{a}_{\mathrm{CLS}}$ representations perform better on the user side.
A possible explanation is that $h^{a}_{\mathrm{TAR}}$ embeddings are more sensitive to sparse interactions and cold-start authors. 
In addition, identity-enhanced representations introduce higher variance across authors, which makes it harder to extract stable long-term user preferences.

\subsection{Analysis Experiments (RQ2)}

\subsubsection{\textbf{Stratified Test.}}

\begin{figure}[t!]
    \vspace{-0.2cm}
    \centering
    \subfloat[AUC\label{fig:vv_ctr_auc}]{
        \includegraphics[width=8.5cm]{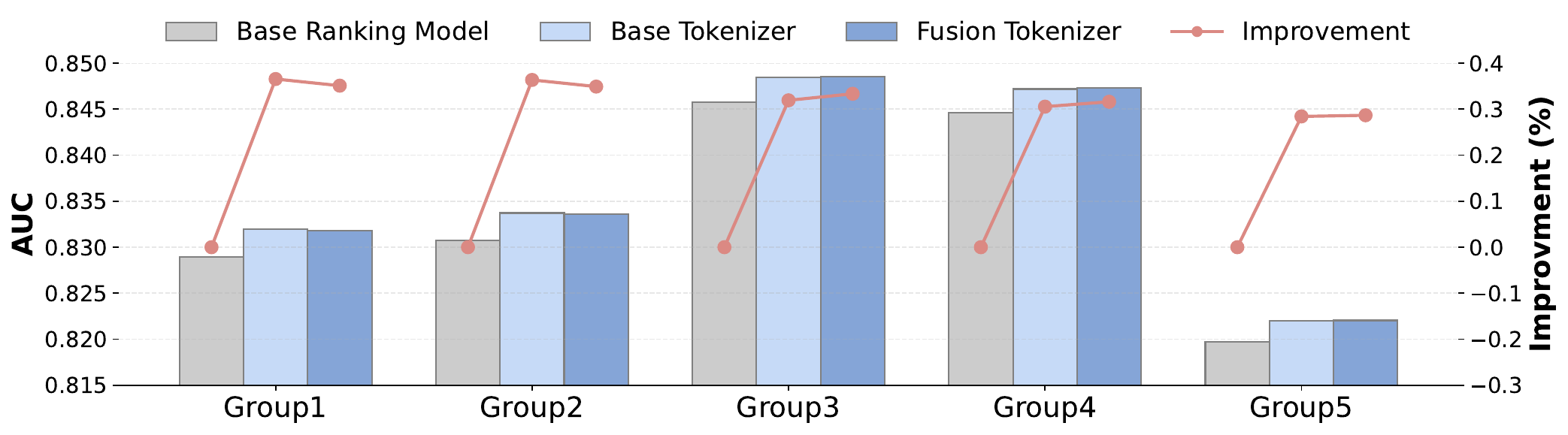}
    }
    \quad
    \subfloat[GAUC\label{fig:vv_ctr_wuauc}]{
        \includegraphics[width=8.5cm]{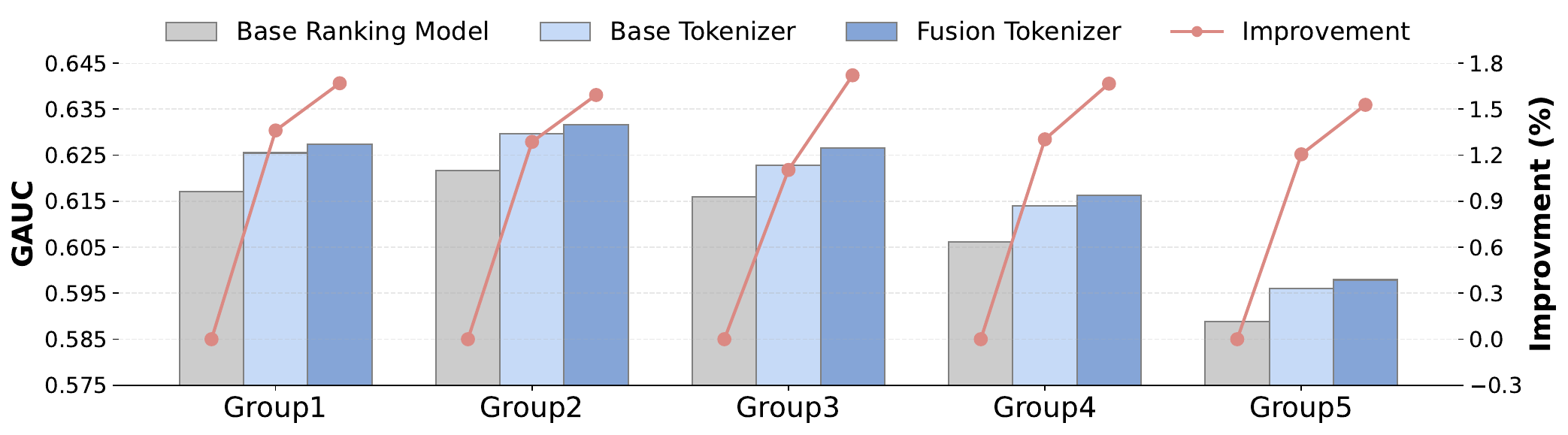}
    }
    \caption{Stratified AUC and GAUC for CTR prediction under different tokenization.}
    \vspace{-0.5cm}
    \label{fig:vv}
\end{figure}

To evaluate how content understanding contributes to author representations across different exposure levels, we conduct a stratified study over five author groups segmented by exposure frequency: $[0,60), [60,100), [100,1000), [1000,10000),$ $ [10000,\infty)$. 
We compare CTR prediction performance of SARM variants using the base tokenizer and the tokenizer fusion, with results shown in Figure~\ref{fig:vv}.
We observe the following:
\begin{itemize}
\item SARM consistently outperforms the base ranking model across all groups, indicating that enhanced content modeling provides a stable source of incremental gains for author representation.

\item Relative improvements are larger in low-exposure groups, suggesting stronger generalization to long-tail authors. 
Importantly, improvements are also observed in high-exposure groups, indicating that gains on long-tail authors are achieved without degrading performance on head authors.

\item Moreover, tokenizer fusion provides consistent AUC gains over the base tokenizer and yields larger improvements in GAUC, suggesting improved personalization under exposure imbalance.
These results align with the stronger gains observed in low-exposure authors.
\end{itemize}

\subsubsection{\textbf{Attention Analysis.}}

\begin{figure*}[t!]
% \vspace{-0.1cm}
\begin{center}
\includegraphics[width=17.0cm]{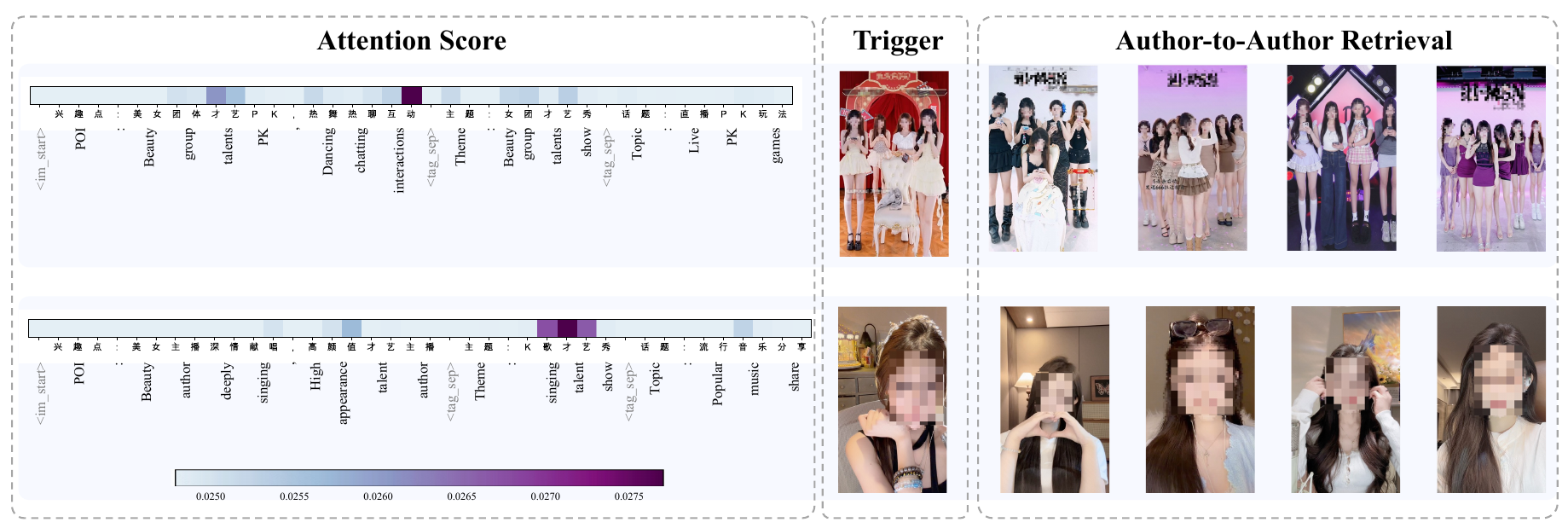}
\vspace{-0.3cm}
\caption{
Attention weight visualization on semantic anchors for a trigger author, with author-to-author retrieval based on semantic similarity.
For clarity, we show three of the six semantic anchor dimensions (POI, Theme, Topic).
}
\vspace{-0.3cm}
\label{fig:case}
\end{center}
\end{figure*}

\begin{table*}[t!]
  \caption{A/B testing results from key pages on online live-streaming service platforms.}
  % \vspace{-0.3cm}
  % \setlength{\tabcolsep}{2.5pt}
  \label{tab:ab}
  \begin{tabular}{c|ccccccc}
    \toprule
    \multirow{2}{*}{\textbf{Applications}} & \multicolumn{3}{c}{\textbf{Core Metrics}}  & \multicolumn{4}{c}{\textbf{Interaction Metrics}} 
    \\ \cmidrule(r){2-4} \cmidrule(r){4-8} & Exposure & Watch Count & Watch Time & Click & Gift & Effective View & Follow \\
    
    \midrule
    
    \multirow{1}{*}{\cellcolor{white}\textbf{Kuaishou}} & {+0.424\%} & {+0.189\%} & {+0.092\%} & {+0.982\%} & {+0.482\%} & {+0.070\%} & {+0.805\%}\\ 

    \midrule

    \multirow{1}{*}{\cellcolor{white}\textbf{Kuaishou Lite}} &  {+1.190\%} & {+0.397\%} & {+0.962\%} & {+0.562\%} & {+1.287\%} & {+0.340\%} & {+0.522\%}\\ 
    
    \bottomrule
  \end{tabular}
  % \vspace{-0.3cm}
\end{table*}

\begin{table}[t!]
  \caption{Improvements over the base model with different MLLMs for semantic anchors.}
  % \vspace{-0.3cm}
  \setlength{\tabcolsep}{2pt}
  \label{tab:mllm}
  \begin{tabular}{c|c|cccc}
    \toprule
    \multicolumn{1}{c|}{\multirow{2.5}{*}{\textbf{MLLM}}} & \multirow{2.5}{*}{\textbf{Benchmark}} & \multicolumn{2}{c}{\textbf{CTR}}  & \multicolumn{2}{c}{\textbf{LVTR}} \\
        \cmidrule(r){3-4} \cmidrule(r){5-6} & & AUC & GAUC & AUC & GAUC \\
    \midrule

    \textbf{Qwen2.5-VL-7B} & 2.308 & +0.19\% & +0.20\% & +0.27\% & +0.22\% \\
    \rowcolor{blue!8}
    \textbf{Qwen3-VL-8B} & \textbf{2.519} & \textbf{+0.24\%} & \textbf{+0.32\%} & \textbf{+0.31\%} & \textbf{+0.38\%} \\
    
    \bottomrule
  \end{tabular}
  % \vspace{-0.3cm}
\end{table}

To qualitatively examine how SARM leverages semantic anchors, we visualize token-level contributions to the identity-aware author embedding $h^{a}_{\mathrm{TAR}}$.
Following~\cite{quantifying}, we approximate token importance by computing the cumulative product of target attention and self-attention scores within SAE.
As shown on the left side of Figure~\ref{fig:case}, the model concentrates on tokens that capture the core semantics of each live stream.
For the \textbf{\textit{Beauty Group}} author, higher weights are assigned to keywords such as \textit{group}, \textit{interaction}, and \textit{talents}, whereas for the \textbf{\textit{Singer}} author, attention shifts toward words like \textit{singer}, \textit{appearance}, and \textit{talents}.
These visualizations illustrate how SARM emphasizes concise semantic cues aligned with the live-streaming content, providing insight into how semantic anchors guide author representation.

\subsubsection{\textbf{Case Study.}}
To examine how semantic anchors organize author representations, we perform Author-to-Author (A2A) retrieval using the [CLS] token. 
Retrieval examples are shown on the right side of Figure~\ref{fig:case}. 
The retrieved authors exhibit strong content similarity to the query author. 
For instance, when the trigger author belongs to the \textbf{\textit{Beauty Group}} or \textbf{\textit{Singer}} category, the system retrieves neighbors with closely related content. 
These examples illustrate how the model clusters authors according to semantic anchor representations within the end-to-end ranking pipeline.

\subsubsection{\textbf{Upgrade Semantic Anchors.}}

To verify that SARM can directly benefit from higher-quality semantic anchors, we evaluate variants built on different fine-tuned MLLM models.
Only the MLLM for generating semantic anchors is replaced, while the ranking model remains unchanged. 
Results in Table~\ref{tab:mllm} show that improvements in anchor quality translate into consistent gains in ranking performance, indicating that SARM can automatically exploit richer textual descriptions without modifying the ranking pipeline. 
Additional analysis is provided in Appendix~\ref{app:mllm}.

\subsubsection{\textbf{Computational Overhead.}}

\begin{table}[t!]
  \caption{Computational cost of the base ranking model and SARM during training and inference.}
  % \vspace{-0.3cm}
  % \setlength{\tabcolsep}{7pt}
  \label{tab:cost}
  \begin{tabular}{l|ccccc}
    \toprule
    \multicolumn{1}{c|}{\multirow{2.5}{*}{\textbf{Model}}} & \multicolumn{3}{c}{\textbf{Training}}  & \multicolumn{2}{c}{\textbf{Inference}} \\
    \cmidrule(r){2-4} \cmidrule(r){5-6} & CPU & GPU & Time/Pass & QPS & Time/Query \\
    
    \midrule

    \textbf{Base} & 48.69\% & 77.54\% & 1.00x & 280.71 & 1.00x \\
    \textbf{SARM} & 51.71\% & 80.29\% & 1.08x & 271.30 & 1.02x \\
     
    \bottomrule
  \end{tabular}
  % \vspace{-0.3cm}

\vspace{-0.3cm}
\end{table}

As shown in Table~\ref{tab:cost}, we compare the computational overhead of SARM with the base ranking model. 
Under identical distributed training settings, SARM introduces only a small increase in offline training cost. 
During online inference, the lightweight deployment design preserves both single-server QPS and per-query latency, with no statistically significant degradation observed. 
These results indicate that SARM maintains deployment efficiency at production scale.

\subsection{Online A/B Test (RQ3)}

We deploy SARM on the key pages of the live-streaming ranking services of the Kuaishou and Kuaishou Lite platforms to assess its impact in production. Results from a two-week online A/B test are summarized in Table~\ref{tab:ab}. SARM consistently improves key interaction and retention metrics on both platforms, while also reducing negative feedback signals. These results suggest that end-to-end semantic modeling leads to measurable gains in user engagement on key pages in real-world deployment.

\section{Conclusion}

In this paper, we present SARM, a new ranking framework for live-streaming recommendation that integrates semantic anchors into end-to-end ranking.
By introducing semantic anchors as ranking-aware units, SARM bridges content understanding with ranking optimization while remaining compatible with industrial deployment constraints. 
Extensive offline evaluation and long-term production A/B tests demonstrate consistent improvements in engagement and commercial metrics. 
SARM is deployed at Kuaishou scale, serving over 400 million users in real-world live-streaming services.

\balance
\bibliographystyle{ACM-Reference-Format}
\bibliography{sample-base-extend.bib}

\newpage

\appendix

\section{More Experiments}

\subsection{Baselines} \label{app:baseline}

The variants we compared with include: 
\begin{itemize}
    \item \textbf{Base Ranking model}: An MMoE-style multi-task learning architecture HoME~\cite{home}, which has been deployed online and serves as the basic of our platform. We also use it as the baseline for the experiment, which incorporates basic functions such as ID embedding, sequence modeling and feature crossing.
    \item \textbf{Tags}: Incorporating tag classification information of authors or living room as features into the ranking model.
    \item \textbf{SIDs}: Incorporating the semantic IDs obtained by processing author's content information through residual quantization to the modal, similar to LARM~\cite{larm}.
    \item \textbf{MLLM Emb}: Incorporating the dense multimodal embeddings of author's content information to the model through tow ways. 
    First, injecting raw multimodal embeddings of the target author, similar to MMBee~\cite{mmbee}. 
    Second, augmenting user's interaction history with multimodal embeddings, similar to SIM~\cite{sim}.
    \item \textbf{SARM}: Incorporating the end-to-end framework we proposed to the ranking model, includes semantic anchor encoding modules on both author and user sides and an auxiliary loss. The corresponding modules are also subjected to ablation.
\end{itemize}

\subsection{\textbf{Different MLLM}} \label{app:mllm}

\begin{figure}[H]
% \vspace{-0.1cm}
\begin{center}
\includegraphics[width=8.5cm]{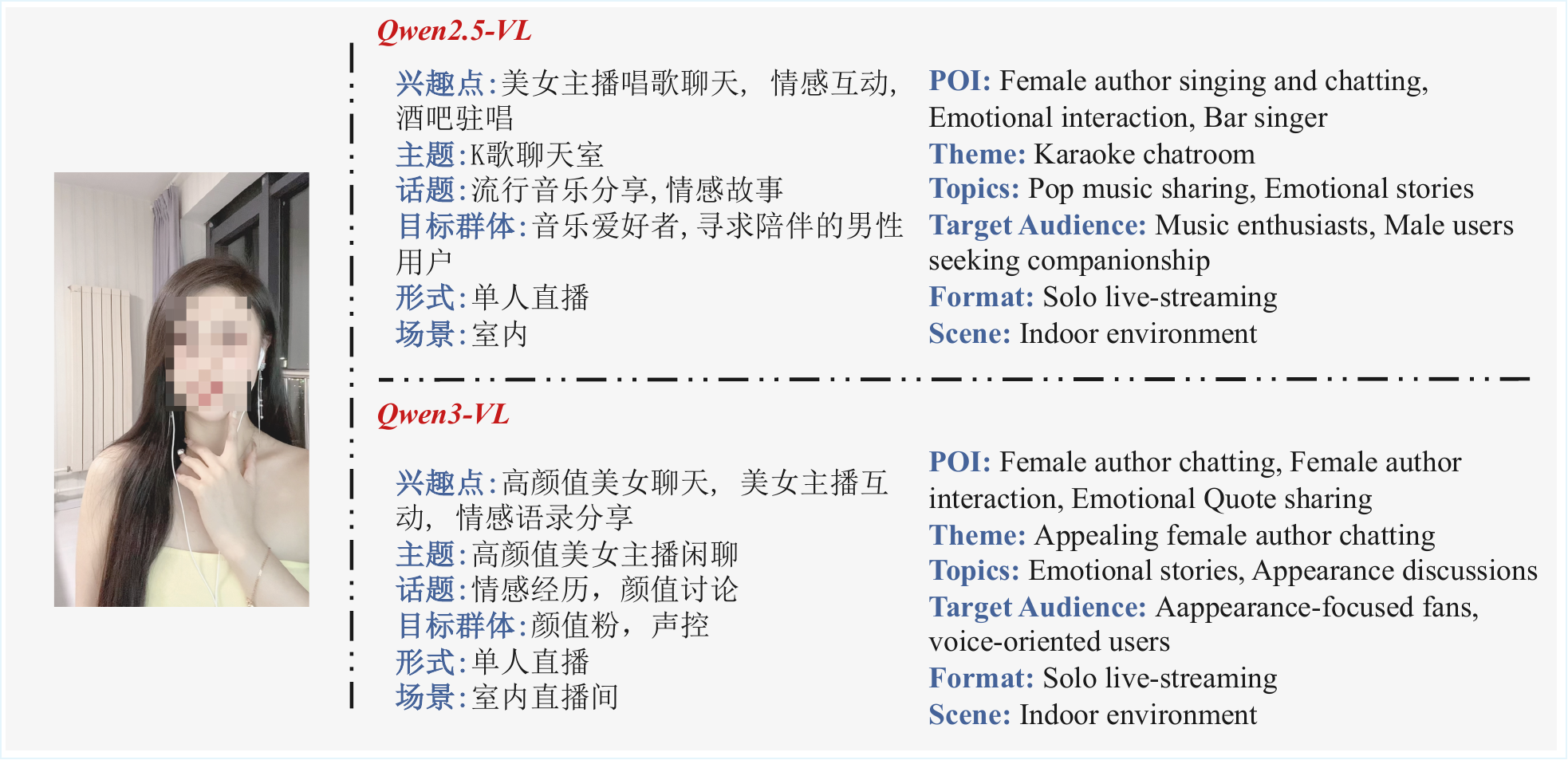}
% \setlength{\abovecaptionskip}{3pt}
% \vspace{-0.3cm}
\caption{Semantic anchor generated by different MLLM.}
% \vspace{-0.5cm}
\label{fig:case_appendix}
\end{center}
\end{figure}

We present the comparison case of semantic anchor generated by different base MLLMs applied to the same live-streaming in Figure \ref{fig:case_appendix}. 
We can observe that the model with poorer performance have illusions in their understanding (\textit{e.g.}, incorrect descriptions \textit{Bar Singer} that do not match the facts), which affects the accuracy of the generated text. Meanwhile, the better model can accurately depict the live-streaming.

In addition, we introduce a lightweight benchmark that evaluates the similarity between generated semantic anchors and live-streaming content. The assessment is conducted across six dimensions, each scored from 0 to 3 (low to high). The results are summarized in Table \ref{tab:mllm}. We observe that stronger models consistently achieve higher scores, demonstrating their superior capability in capturing and representing live-streaming content.

\begin{table}[H]
  \caption{Benchmark of similarity between semantic anchors and live-streaming content.}
  % \vspace{-0.3cm}
  \setlength{\tabcolsep}{4pt}
  \label{tab:mllm}
  \begin{tabular}{c|cccccc}
    \toprule
    \textbf{MLLM} & POI & Theme & Topic & Target & Format & Scene \\

    \midrule

    \textbf{Qwen2.5-VL} & 2.157 & 2.221 & 2.197 & 2.344 & 2.515 & 2.415 \\
    \textbf{Qwen3.0-VL} & 2.468 & 2.559 & 2.418 & 2.388 & 2.773 & 2.505 \\
    
    \bottomrule
  \end{tabular}
  % \vspace{-0.3cm}
\end{table}

\subsection{\textbf{Insertion Position}} \label{app:position}

We evaluate different insertion positions of the gated fusion module, with results shown in Table \ref{tab:position}. We take inserting the module after the first self-attention layer as the baseline, and report other variants as relative improvements over this baseline. 
The results indicate that applying gated fusion after every layer yields the best performance. We attribute this to the fact that such a design continuously injects domain-enhanced signals throughout the network, leading to more robust and information-rich content representations.

\begin{table}[ht!]
  \caption{Comparison of insertion positions of gated fusion.}
  % \vspace{-0.3cm}
  % \setlength{\tabcolsep}{7pt}
  \label{tab:position}
  \begin{tabular}{l|cccc}
    \toprule
    \multicolumn{1}{c|}{\multirow{2.5}{*}{\textbf{Positions}}} & \multicolumn{2}{c}{\textbf{CTR}}  & \multicolumn{2}{c}{\textbf{LVTR}} \\
    \cmidrule(r){2-3} \cmidrule(r){4-5} & AUC & GAUC & AUC & GAUC \\
    
    \midrule

    \textbf{Behind Layer 0} & - & - & - & - \\
    \textbf{Behind Layer 0 \& 1} & +0.02\% & +0.02\% & -0.02\% & +0.12\% \\
    \textbf{Behind Layer 0 \& 1 \& 2} & +0.06\% & -0.02\% & +0.03\% & +0.14\% \\
    
    \bottomrule
  \end{tabular}
  % \vspace{-0.3cm}
\end{table}

\section{Prompt Template} \label{app:template}

We present the prompt template for semantic anchor generation as bellow:

\begin{tcolorbox}[
    float*=ht!,
    title=Prompt for Semantic Anchor Generation,
    width=\textwidth, 
    colframe=gray!60
]
    % \small 
    \textbf{[Author Profile] [Visual KeyFrames] [Audio Transcription] [Text Comments]} \newline
     
    The above describes the streamer’s basic information.
You are a content understanding expert who has watched a large number of Kuaishou live-streamings. Based on a full understanding of the live-streaming content, generate summary tags step by step. These tags will be used for live-streaming distribution, recommendation, and search. Each step’s output must be explicitly included in the results. \newline

\textbf{Step 1: Live-Streaming Content Understanding \& Summary}

\textbf{Task:} Summarize the live-streaming content, identify the main theme, and infer the author’s motivation. \newline

\textbf{Step 2: Candidate Tag Generation \& Reasoning}

\textbf{Task:}  Based on the content understanding, generate the following types of tags:

\textbf{1.Point of Interest:} Key elements that attract users, up to 3 (e.g., “Tibetan beauty chatting in dialect,” “Middle-aged man angrily reviewing cars”).

\textbf{2.Theme:} A short phrase summarizing the live-streaming, up to 1 (e.g., “KOF ’97 Classic Revival,” “College Entrance Exam Application Guide”).

\textbf{3.Topic:} Discussable topics involved in the stream, up to 2 (e.g., “Stock market trends,” “Tangshan local life”).

\textbf{4.Target Audience:}  User groups suitable for recommendation, up to 2 (e.g., “Stock investors,” “Parents of second children”).

\textbf{5. Format:} Presentation format of the live-streaming (e.g., “Single-author live-streaming,” “Multi-author live-streaming,” “Outdoor live-streaming,” “Unattended live-streaming”).

\textbf{6. Scene:} Specific scenarios involved (e.g., “Home interior,” “Public gym,” “Virtual studio,” “Mobile screen”). \newline

\textbf{Step 3: Classification \& Formatted Output}

\textbf{Task:}  Organize the tags from Step 2 into a JSON format by dimension name + value list (even if empty or containing only one item), strictly as:

{\centering \textbf{The Format of Semantic Anchor} \par}

Each field must be a list of phrases.
    
\end{tcolorbox}

\end{document}